\documentclass[twocolumn,
aps,nofootinbib,showpacs,showkeys,preprint
tightenlines
] {revtex4-1}

\usepackage{epsf,epsfig,subfigure,axodraw,graphicx,amsmath,amssymb}
\usepackage{color}
\newcommand{\dis}[1]{\begin{equation}\begin{split}#1\end{split}\end{equation}}

\newcommand{\ie}{{\it i.e.}\ }
\newcommand{\etal}{{\it et al.}\ }

\def\sech{{\rm sech}}

\def\mt2{M_{T2}}
\def\trialmt2{M_{T2}(\chi)}

\def\mthetat2{M_{\theta T2}}
\def\trialmthetat2{M_{\theta T2}(\chi)}

\def\mpit2{M_{\pi T2}}
\def\trialmpit2{M_{\pi T2}(\chi)}

\def\p0{{\bf p}_0}

\def\ie{{\it i.e.}~}
\def\etal{{\it et al.}}

\begin{document}

\title{\Large\bf Cosmological constant is probably adjustable in brane worlds}

\author{Jihn E. Kim\email{jekim@ctp.snu.ac.kr}}
\affiliation{ Department of Physics and Astronomy and Center for Theoretical Physics, Seoul National University, Seoul 151-747, Korea
 }
\begin{abstract}
In the brane world scenario, possessing the self-tuning property of the cosmological constant, we study the probability amplitude for an initial state of the Universe to transform to final states with some curvatures. In the Kim, Kyae and Lee model, there exists a finite range of parameters such that the transition amplitude to a near flat universe is exponentially dominated by the smallness of curvature, $\gtrsim e^{{\rm positive~ number}/\,\overline{\Lambda}^2}$, i.e. is dominated by an almost flat universe of $|\overline{\Lambda}|<\epsilon$.
\end{abstract}


\maketitle

%%%%%%%%%%%%%%%%%%%%%%%%%%%%%%%%%%%%%%%%%%%%%%%%%%%%%%%%%%%%
\section{Motivation and strategy}\label{sec:Introduction}

There are three cases where the cosmological constant(CC) appears explicitly in the evolution of the Universe. The most well-known problem is the fine-tuning problem of the present CC \cite{WeinRMP}. The second case is a huge CC,  $\Lambda_{\rm infl}$, needed in inflationary models \cite{Inflation}. $\Lambda_{\rm infl}$ here is somewhat below the Planck scale  $\Lambda_{\rm infl}<(10^{-6}M_P)^2$, or $V_{\rm vac}<(10^{-3}M_P)^4$ in terms of the vacuum energy. The third CC problem is to understand why the current vacuum energy, observed in 1998, is so tiny at the order of $(2\times 10^{-3}{\rm eV})^4$ \cite{SNcc98}. Among these, the most difficult CC problem is the first one, the fine-tuning problem of $V_{\rm vac}$ compared to the energy scale in consideration.

In quantum theory, paths of a field configuration are integrated over to obtain the transition amplitude from an initial state to a final state. At present, there does not exists a well defined field theory of quantum gravity. Nevertheless, (in the hope of finding a hint) one may guess that from the path integral approach some gravity effects can be attainable, in particular for obtaining the classical terms. This may be justified for situations where the relevant scale of physics is far below the reduced Planck mass $M_P=2.44\times 10^{18}$ GeV. For the vacuum transition amplitude, it has been argued that this truncation to classical terms might be useful \cite{Hawking84}. With this philosophy, Baum and Hawking used the Euclidian action for the probability amplitude from an initial state $|I\rangle$ to a final state $|\overline{\Lambda}_f \rangle$  \cite{Hawking84},
\dis{
\langle\overline{\Lambda}_f |I \rangle \propto \int d[g]~ e^{-I_E[g]}
\label{eq:ProbinEucl}
}
where the Euclidian action $I_E$ is calculated with the wave function specified by the metric $g$ and the initial state $|I\rangle$.

In 4 dimensions(4D), the CC is a fixed parameter and we can infer naively that  $e^{-I_E[g]}$ diverges for a small $\overline{\Lambda}$ as $e^{{\rm (positive~constant)}/\,\overline{\Lambda}}$, leading to the probability amplitude dominated by a large positive exponent for $\overline{\Lambda}=0^+$ \cite{Hawking84}. The detailed reason for ${{\rm (positive~constant)}/\,\overline{\Lambda}}$ in the exponent in the Euclidian space comes from the Euclidian action $-I_E=\int d^4x_E(\frac12 R_{E} - V_0)$ where $V_0$ is the vacuum energy. The Einstein equation relates the curvature $\overline{\Lambda}$ to the vacuum energy, $\overline{\Lambda}=V_0$. In the maximally symmetric space, the Einstein tensor is simply related to the curvature by $R_{\mu\nu}=\overline{\Lambda} g_{\mu\nu}, R_{E}=4\overline{\Lambda}$ and hence $\frac12 R-V_0=\overline{\Lambda}$. So, we notice that the contribution to the Euclidian space integral from $\frac12 R_{E}$ for the de Sitter (dS) space case is positive, dominating the negative contribution of the particle physics contribution, $-V_0$. Hawking used this fact to state, ``Cosmological constant is probably zero."

However, one should notice that this argument of choosing $\overline{\Lambda}=0^+$ is based on the fact that the Euclidian space action $-I_E$ is dominated by one negative power of $\overline{\Lambda}$ as $1/\,\overline{\Lambda}$. If it is dominated by two negative powers of $\overline{\Lambda}$ as $1/\,\overline{\Lambda}^2$, which is not unreasonable in self-tuning models, the argument changes completely to the case that $\overline{\Lambda}=0$ is approached from the Anti-dS(AdS) space side. In fact, we will show in this paper that this behavior is happening in the self-tuning model of Kim, Kyae and Lee(KKL) \cite{KKL00,KKL01}.

It is well known that {\it in 4D} the first CC problem is a fine-tuning problem and it does not have a solution \cite{WeinRMP}. Furthermore, one needs a large vacuum energy for inflation in the very early universe, exiting to a final state with a negligible vacuum energy. To understand the first CC problem, therefore, one may go beyond 4D with extra space dimensions. For example, with one extra space added, \ie in 5D, the amplitude $\langle\overline{\Lambda}_f |I \rangle $  may not be proportional to a delta function and then one can literally talk about the probability of  transition from $|I \rangle$ to $|\overline{\Lambda}_f\rangle$. This has been a reason for the renewed CC study in 5D at the turn of the century \cite{Arkani00,Forste00,Csaki00,KKL00,KKL01}, under the name of {\it self-tuning} solutions. In a typical Randall-Sundrum model of type II (RS-II) \cite{RSII}, the 5D bulk CC $\Lambda_b$ must be fine-tuned to a certain value of the brane tension $\Lambda_1$. This requires a fine-tuning and does not lead to a solution even in 5D \cite{Forste00,Csaki00}. The only existing {\it self-tuning} model in 5D is the one presented in Ref. \cite{KKL00}.
Here, we do not worry about how such a KKL term, starting from a fundamental Lagrangian, can be written below the 5D  fundamental scale scale.\footnote{At the field theory level, one may try to obtain an effective theory from a fundamental Lagrangian along the line of Wilson and Weinberg \cite{efftheory}. At present, we are far from adopting this strategy.} We try to investigate just the effect of the self-tuning property of some theory.

\begin{figure}[t!]
 \begin{center}
 \epsfig{figure=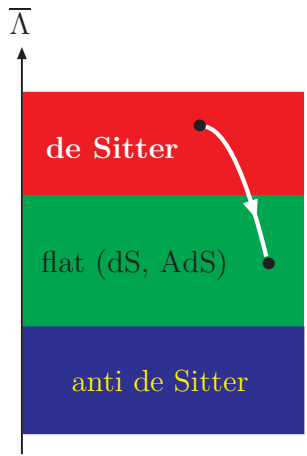,width=4cm,height=6cm,angle=0}
\caption{ The region $|\Lambda_1|<\sqrt{-6\Lambda_b}$  allows a finite range of parameters for the 4D flat space.  }
\label{fig:Blowingup}
\end{center}
\end{figure}

The self-tuning model of Ref.  \cite{KKL00} shows a certain nice behavior. The flat solution is possible for some finite range of the brane tension $\Lambda_1$ compared to the 5D AdS bulk cosmological constant $\Lambda_b$ \cite{Kim03}.  This situation is depicted in Fig. \ref{fig:Blowingup}. If the brane tension $\Lambda_1$ is in the red color region of Fig. \ref{fig:Blowingup}, only the dS space solution is allowed. The brane tension $\Lambda_1$ is determined by dynamics of the observable sector at the brane. So, this can be used as a model for inflation at the brane. Then, the inflationary phenomenology is described by physics at the brane.

The exit from the inflationary period always accompanies a huge change of the vacuum energy. So, after the inflation $\Lambda_1$ changes drastically to be placed in the green region of Fig. \ref{fig:Blowingup} where dS, flat and AdS solutions are possible. In this paper, we will begin physics from a new $\Lambda_1$ placed in the green region of  Fig. \ref{fig:Blowingup}, and then estimate the transition amplitude to  flat, dS and AdS solutions. This setup does not have any ambiguity for the initial state condition after inflation. We choose it to be the eigenstate of the Hamiltonian, $|\Lambda_1\rangle$, in the green region.

So, our strategy to solve all CC problems is using a self-tuning solution containing a {\it flat-space-forbidden region}. At present, we are not sure about that this property survives in all self-tuning models. The first CC problem is solved just by showing a diverging probability of (\ref{eq:ProbinEucl}) in the flat-space-allowed region, which is our main calculation in this paper. The second CC problem is solved just by assuming the Universe starting from a {\it flat-space-forbidden region} in self-tuning models. The third CC problem is attributed to a small difference of the potential energy in the brane as done in Ref. \cite{quint03}, which is not repeated in this paper.

%%%%%%%%%%%%%%%%%%%%%%%%%%%%%%%%%%%%%%%%%%%%%%%%%%%%%%%%%%%%%
\section{Eigenstates of the Universe}\label{sec:Eigenstates}

In the one-brane world scenario with a ${\bf Z}_2$ symmetry, we assume that there exists the brane tension $\Lambda_1$ at the $y=0$ brane and the bulk ($y\ne 0$) cosmological constant $\Lambda_b$ as in the RS-II model \cite{RSII}. There exists a ${\bf Z}_2$ symmetry: $y\to -y$. In addition, we assume that there exist(s) additional fields such that self-tuning solutions are possible as done in terms of three index antisymmetric tensor gauge field $A_{MNP}$ and its field strength $H_{MNPQ}$ in \cite{KKL00,KKL01}. For a given $\Lambda_1$, it is possible to have multiple solutions. For simplicity of discussion, we restrict to the maximally symmetric spaces of  flat, dS, and AdS solutions. In quantum mechanics, our initial state vector is,
\dis{
|I\rangle =|\Lambda_1\rangle .\label{eq:Lambdaonestate}
}
The bulk cosmological constant $\Lambda_b$ is a God-given number in the theory and is not a changeable parameter. However, $\Lambda_1$ is treated as a changeable parameter due to physics at the $y=0$ brane. A sudden change of $\Lambda_1$ at the brane is a new state chosen by a filtering process in quantum mechanics. We assume that the complete set of maximally symmetric states in the bulk of the Universe covers flat, dS and AdS states
\dis{
|g_{MN}, {\rm bulk~ fields}\rangle= |\overline{\Lambda},h\rangle\label{eq:CompleteU}
}
where $h$ denotes integration constants in the bulk in self-tuning models and the 4D curvature $\overline{\Lambda}$ is assumed to be independent of $y$.
Consider an initial state $|\Lambda_1\rangle$. It is immediately related to a nearby bulk state which we again call $|\Lambda_1\rangle$. Technically, this corresponds to the boundary conditions of the bulk field values at $y=0^+$. This properly-connected bulk state corresponds to the brane with the tension $\Lambda_1$. This boundary condition does not choose the curvature $\overline{\Lambda}$ uniquely, and here we talk about the probability amplitude to choose a $\overline{\Lambda}$. Namely, we are interested in the probability amplitude $C(\overline{\Lambda}, h_i)$ obtaining a $\overline{\Lambda}$ universe, starting from the brane tension $\Lambda_1$,
\dis{
|\Lambda_1\rangle = \int [dh]\int [d\overline{\Lambda}] C(\overline{\Lambda}, h) | \overline{\Lambda}, h\rangle
}
where
\dis{
 C(\overline{\Lambda}, h)=\langle \overline{\Lambda}, h |\Lambda_1\rangle .
}
The probability amplitude is the form given in Eq. (\ref{eq:ProbinEucl}), and the relevant factor is
\dis{
\int [d\Psi] [d\overline{\Lambda}][dh]C(\overline{\Lambda}, h)\propto\int [d\Psi] [d\overline{\Lambda}][dh]~ e^{-I_E[\Psi, \overline{\Lambda}, h]}
\label{eq:ProbAmpl}
}
where $\Psi$ is the warp factor introduced in the metric.

%%%%%%%%%%%%%%%%%%%%%%%%%%%%%%%%%%%%%%%%%%%%%%%%%%%%%%%%%%%%%
\section{Solutions of the self-tuning KKL model}\label{sec:KKLsolution}
To show some effect of self-tuning models in the CC problem, we take up the known example \cite{KKL00} whose key points are summarized below.
In the 5D Euclidian space with one 3-brane at $y=0$, the KKL model with the three index antisymmetric-tensor gauge field $A_{MNP}$ is given by \cite{KKL00,KKL01},
\dis{
-I_E &=\int d^5x_E \sqrt{g_{(5)}}\left(\frac12 R_{(5)}-\frac{2\cdot 4!}{H^2}-\Lambda_b -\Lambda_1\delta(y) \right)\\
&= \int dy\int d^4x_E  \Big\{-\Psi^4\Lambda_1\delta(y)+\frac12 R\Psi^2 \\
&\quad  + 4\Psi^3\Psi''+6\Psi^2(\Psi')^2 +\frac{2\cdot 4!\Psi^4}{H^2}-\Psi^4\Lambda_b
 \Big\} \label{eq:ActionKKL}
 }
where the 5-dimensional fundamental scale $M$ is set to 1, $H_{MNPQ}$ is the field strength of $A_{MNP}$, $'$ denotes the derivative with respect to $y$ and the metric is given by
\begin{equation}
ds_E^2=\Psi^2(y)g_{\mu\nu}dx_E^\mu dx_E^\nu +dy^2.\label{eq:5DMetric}
\end{equation}
The solution satisfying the field equation and the Bianchi identity has been given in \cite{KKL00,KKL01} for
\dis{
H_{\mu\nu\rho\sigma}=\sqrt{g}\frac{\epsilon^{(E)}_{\mu\nu\rho\sigma}}{n(y)}, ~ H_{5\nu\rho\sigma}=0,~ \epsilon^{(E)}_{0123}= \epsilon_{(E)}^{0123}=1\label{eq:Hsolution}
}
with $n(y)=h^{-1}\sqrt{g}=h^{-1}\Psi^4(y)$, where the Greek indices $\mu,\nu,\cdots$ run over the 4D space. Then, $H^2$ given by $g^{AM}g^{BN}g^{CP}g^{DQ}H_{ABCD}H_{MNPQ}$ becomes
$H^2=4!\Psi^{-8}h$ and hence $1/{H^2}={\Psi^{8}}/(4!h)$, where $h$ is the integration constant (with $ h>0$) determined by the solution of the differential equations with appropriate boundary conditions.
Two relevant bulk Einstein equations are \cite{KKL01}
\dis{
&-3\frac{\overline{\Lambda}}{\Psi^2}+3\left(\frac{\Psi'}{\Psi}\right)^2
+3\frac{\Psi''}{\Psi}=- \Lambda_b - 2\cdot 4!\left(\frac{3}{H^2}\right) \\
&-\frac{6}{\Psi^2}\overline{\Lambda} +6\left(\frac{\Psi'}{\Psi}\right)^2
=-\Lambda_b -2\cdot 4!\left( \frac{1}{H^2}\right).\label{eq:Einstein}
}
Or, keeping the brane $\Lambda_1$ term also, we have
\dis{
\frac14 Y'' = 3{\overline{\Lambda}}\sqrt{Y}+\frac{2m^2k^2}{3}Y- \frac{8}{3h}Y^3-\frac{\Lambda_1}{3}\delta(y)Y\label{eq:Ydef}
}
plus one of (\ref{eq:Einstein}), where $Y=\Psi^4, \Lambda_b=-m^2k^2$.
From Eq. (\ref{eq:Ydef}), we find that $\Psi(0^+)$ at $y=0^+$ becomes
\dis{
\Psi'(0^+)=-\frac{\Lambda_1}{6}A,\quad A=\Psi(0^+).\label{eq:BC0}
}
From the flat-space solution \cite{KKL00}, $Y$ of  the form $Y=A[\sech(ky+c_0)+c(y)]$ can be tried for the nearby dS and AdS solutions
\dis{
A&k^2\sech-2Ak^2\sech^3+Ac'' =12\frac{\overline{\Lambda}}{\sqrt{Y}} +A\frac{8m^2k^2}{3}\sech\\
&\quad - \frac{32A^3}{3h}(\sech^3+3\sech^2c+3\sech c^2+c^3)
}
where the argument of $\sech$ is $(ky+c_0)$. So, for the flat, dS, and AdS cases, we determine
\dis{
m^2=\frac38,\quad \frac{16A^2}{3h}=k^2.\label{eq:hmfree}
}

%%%%%%%%%%%%%%%%%%%%%%%%%%%%%%%%%%%%%%%%%%%%%%%%%%%%%%%%%%%%%%%%%%%%%%%%%
\section{The probability amplitude}

In calculating the probability amplitude in 4D with $H_{\mu\nu\rho\sigma}$, there have been some arguments which value of the $H^2$ term must be used in the action integral. For example, with the $\Lambda$ and $R$ terms,  Duff shows that substituting an ansatz into an action and varying that action does not yield the same result as substituting an ansatz into the equation  of motion \cite{Duff89}. To obtain the same result, Duff considers a surface term $\epsilon^{\mu\nu\rho\sigma}H_{\mu\nu\rho\sigma}$ \cite{Aurelia80}, which does not change the equation of motion. If a surface term value is added under Duff's condition, he shows that it amounts to changing the sign of $H^2$ term in the action integral \cite{Duff89} and concluded that the probability amplitude for $\overline{\Lambda}=0$ is least probable with $H_{\mu\nu\rho\sigma}$. Recently, however, Wu obtained the opposite result from that of Duff \cite{Wu08}. This example shows that we need a care when surface terms are considered.

In our case also, we may consider a surface term (due to $x$-independent $\Psi^8/H^2$), in the language of the first order formalism of Ref. \cite{Aurelia80},
$$
-I_{E}\supset\int dy\int d^4x_E\left\{ \rho\frac{2\Psi^{8}}{H^2} \epsilon^{\mu\nu\rho\sigma}H_{\mu\nu\rho\sigma} -\frac{\Psi^{4}}{2} \rho^2\right\},
$$
and so we need a care. When a surface term is considered, it involves instanton solutions and hence the vacuum structure must be properly defined as done in QCD.
In our case at hand, we must consider an $\varphi$-vacua, similar to the $\theta$-vacua of QCD. For the simplicity of the algebra, we show the $\varphi$-vacuum with the $H^2$ term. [For $1/H^2$ term, the argument goes in parallel.]

For two antisymmetric indices from $\mu,\nu,\rho,$ and $\sigma$, there are
six ($_4C_2=6$) independent second rank antisymmetric gauge
functions, for which $A_{\mu\nu\rho}$ transforms as
\dis{
A_{\mu\nu\rho}\to A_{\mu\nu\rho}-\partial_\mu \Lambda_{\nu\rho}-\partial_\nu \Lambda_{\rho\mu}-\partial_\rho \Lambda_{\mu\nu}~.
}
The gauge symmetry of the instanton solution is given by any six directions of $\Lambda_{\mu\nu}$, three for the instanton ($\Lambda_{ij}$) and three for the anti-instanton ($\Lambda_{0i}$) and the instanton action is $\int d^5x \partial_y\epsilon^{\mu\nu\rho\sigma} H_{\mu\nu\rho\sigma}$. Namely, there exist maps of
\dis{
S_3\to S_3.
}

For the surface term $\int d^4x H_{\mu\nu\rho\sigma}$ to be finite, $H_{\mu\nu\rho\sigma}$ should tend to $r^{-4}$ for a large $r$. In the bulk, it arises from a 2D curl in 5D,
\dis{
&\int dydx^0 ~ \vec{\nabla}\times \vec{A}=\int d\vec{s}\cdot \vec{A}=\int dydx^0 [\partial_y {\rm A}_0 -\partial_0 {\rm A}_5 ]\\
&=\int dydx^0 \partial_y {\rm A}_0 =\int dx^0  {\rm A}_0 =\int d^4x H_{\mu\nu\rho\sigma}
}
where
\dis{
{\rm A}_0 &= \int d^3{\bf x}~\partial_0 A_{ijk}= \int d^3{\bf x}~H_{\mu\nu\rho\sigma}\\
{\rm A}_5 &= \int d^3{\bf x}~\partial_5 A_{ijk}= \int d^3{\bf x}~H_{5ijk}
}
So a gauge invariant instanton of size $\rho$ located at $x=x_0$ takes the form
\dis{
A_{\alpha\beta\gamma}\propto \frac{\epsilon_{\alpha\beta\gamma\mu}x^\mu}{(r^2+\rho^2)^2},\quad r=|x-x_0|
}
so that $A_{\mu\nu\rho}$ is proportional to $r^{-3}$, and $H_{\mu\nu\rho\sigma}$is proportional to $r^{-4}$.
The gauge symmetry of the instanton solution is given by any six directions of $\Lambda_{\mu\nu}$, three for the instanton ($\Lambda_{ij}$) and three for the anti-instanton ($\Lambda_{0i}$) and the instanton action is $\int d^5x \partial_y\epsilon^{\mu\nu\rho\sigma} H_{\mu\nu\rho\sigma}$.
In nonabelian gauge theories, there are many possible gauge configurations such that the irreducible instanton solution give many possible integers for the Pontryagin index. On the other hand, in our case at hand $H_{\mu\nu\rho\sigma}$ instanton gives only $\pm 1$ for the Pontryagin index.

Now, let us construct a gauge invariant $\varphi$-vacuum,
\dis{
|\varphi\rangle = \sum_{n=-\infty}^{+\infty}|n\rangle e^{in\varphi}
}
In the $\varphi$-vacuum let us integrate out the $H^2$ field first. Then, what Duff chose is equivalent to $\varphi=\pi$ and what Wu chose is equivalent to  $\varphi=0$. As in the $\theta$-vacuum of QCD, any value of $\varphi$ is a possibility.

As commented above, this $\varphi$-vacuum can be defined also with the $1/H^2$ term. We calculate the action integral for $\varphi=0$ [with the original sign of the $1/H^2$ term] and $\pi$ [with the sign of the $1/H^2$ term flipped]. For any other values of $\varphi$, the value of the action integral is between them.

In the maximally symmetric space with a constant $\overline{\Lambda}$, the Euclidian Ricci scalar is $R=4\overline{\Lambda}$, and hence we consider the quantity in the curly bracket of Eq. (\ref{eq:ActionKKL}), neglecting the surface term but with the plus and minus signs of the $1/H^2$ term, as $
2\overline{\Lambda}\Psi^2+4\Psi^3\Psi''+6\Psi^2(\Psi')^2
-\Psi^4\Lambda_b \pm 2\frac{ 4!}{H^2}-\frac{\Lambda_1}{2}\delta(y)$,
which is integrated over the 4D Euclidian space, from 0 to $a_{\rm max}$, and then over the warped 5th space, from  0 to $d_m$. $a$ is the length parameter in the maximally symmetric 4D space,  ${a}^2\propto 1/ |\overline{\Lambda}|$, $d_m$ is the $y$-space size. The integration region in the 4D space is up to $a_{\rm max}$ where $a_{\rm max}=a\xi_{\rm max}$ with $\xi_{\rm max}=1$ and $\infty$ for the dS and AdS spaces, respectively. For the flat space, the curvature part of the 4D integral is 0 but we keep the particle physics part as if it has a finite $\overline{\Lambda}$, perform the $y$ integration, and finally take the limit $\overline{\Lambda}\to 0$. The $4$-dimensional dS (flat) volume is $2\pi^2a_{\rm dS}^4$ with the metric  $ds^2= a_{\rm dS}^2 [dr^2/(1-kr^2)+r^2 d\Omega_3^2]$ with $k=1(0)$. The $4$-dimensional AdS volume diverges for whatever value of $|\overline{\Lambda}|$, but we take out the scale factor $a_{\rm AdS}^4$ to compare the results of integration for different $|\overline{\Lambda}|$'s. To compare the results for different $\overline{\Lambda}$'s in the AdS, we Weyl-transform the metric to $ds^2= a_{\rm AdS}^2 f^2(\eta)(d\eta^2+\eta^2 d \Omega_3^2)$ where $f(\eta)=r/\eta =2/(1-\eta^2)$. Then, the 4D AdS volume is regularized, as shown in Appendix, to $(12{\pi}/|\overline{\Lambda}|)^2 \Gamma(-3)/\Gamma(-1)$ where $a^4_{\rm AdS} \propto 1/|\overline{\Lambda}|^2$.   Therefore, performing the action integral over $2\int_{0}^{d_m} dy\int_0^{a^2_{\rm max}} \pi^2r^2dr^2 $, we obtain
\dis{
-I_E\propto \left\{
\begin{array}{ll}
{M_P^4a_{\rm dS}^4}\times O(k y_H),& \  {\rm dS}\\
{M_P^4a_{\rm dS}^4}\times O(\infty)\times O(\frac{k}{|\overline{\Lambda}|^{1/2}}),
&\  {\rm flat}\\
{M_P^4a^4_{\rm AdS}} \times O(\infty)\times O(ky_P), &\ {\rm AdS}
\end{array}\right. \label{eq:ActionResult}
}
where the last factor denotes the result of the $y$ integration. All expressions are proportional to $M_P^4/\, \overline{\Lambda}^2$. In the KKL model, $y_H$ is the horizon distance in the 5th direction and $y_P$ is the first $y$ value where $\Psi'=0$ \cite{KKL00}. The $y$ integration can give a positive or a negative result, depending on the value of $\Lambda_1$.

\begin{figure}[t!]
 \begin{center}
 \epsfig{figure=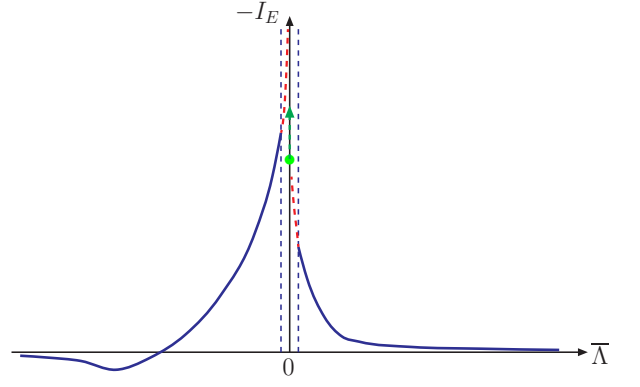,width=8cm,height=5cm,angle=0}
\caption{ The Euclidian space action in the KKL model with an exaggerated dS part. }
\label{fig:LfromAdS}
\end{center}
\end{figure}

To have a positive result, including the brane tension contribution, the following condition must be satisfied
\dis{
\tanh(c_0)\sech^2(c_0)\leq \frac{k}{3} F(c_0/k,d_m)\label{eq:positive}
}
where $F(c_0/k,d_m)$ is the result of the integration. If the condition (\ref{eq:positive}) is not satisfied, $\overline{\Lambda}=0$ would be least probable.  In the integral, the following one is the dominant contribution to the $1/\, \overline{\Lambda}^2$ term,
\dis{
F&(c_0/k,d_m)\simeq \int_0^{d_m} dy \left(\frac{14}{8}\sech\pm\frac{22}{8}\sech^3\right)\label{eq:intdef}
}
where $\pm$ corresponds to $\varphi=\pi,0$, respectively, and the arguments of $\sech$ is $(ky+c_0)$. $c_0$ is the argument at $y=0$. The integral is a function $G(y)$, and $F$ is $G(d_m)-G(0)$ which can be made positive. Here,  $G(\infty)$ has a simple expression,
\dis{
\frac{14}{8k}\sin^{-1}(\tanh)
+\frac{22}{8}\frac{1}{2k}\sinh\sech^2+\frac{22}{8}\frac{1}{2k}\tan^{-1}(\sinh)
\nonumber
}
which is $\frac{9}{2k}\frac{\pi}{2}$ and $\frac{3}{8k}\frac{\pi}{2}$, respectively, for $\varphi=\pi$ and 0. With the inequality (\ref{eq:positive}) satisfied, a schematic behavior of $-I_E$ is shown in Fig. \ref{fig:LfromAdS}, with an exaggerated dS part.
%%%%%%%%%%%%%%%%%%%%%%%%%%%%%%%%%%%%%%%%%%%%%%%%%%%%%%%%%%%%%%%%%%%%%%%
%%%%%%%%%%%%%%%%%%%%%%%%%%%%%%%%%%%%%%%%%%%%%%%%%%%%%%%%%%%%%%%%%%%%%%%%%

We are interested in the parameter space where Eq. (\ref{eq:positive}) is satisfied, where $e^{-I_E}$ behaves as
\dis{
e^{-I_E}\propto \exp\left[ \frac{\rm positive~ number}{\overline{\Lambda}^{2}}\right]\label{eq:PrAmpEst}
}
from which we note:\\
$\Large \bullet$ The exponential factor diverges at least as $e^{{\rm | const.|}/\,\overline{\Lambda}^2}$,\\
\indent which is equivalent to almost going to the flat space\\
\indent  regime.\\
$\Large \bullet$  The dominant contribution to the divergence is\\
\indent  from the AdS vacuum energy $\Lambda_b$, and a somewhat\\
\indent  smaller bulk matter($H_{MNPQ}$) contribution has the\\
\indent  same sign as that of the AdS energy contribution.\\
$\Large \bullet$ Near $\overline{\Lambda}=0$, the AdS space is preferred. But, slightly\\
\indent  outside $\overline{\Lambda}=0$, the dS solution is preferred.

%%%%%%%%%%%%%%%%%%%%%%%%%%%%%%%%%%%%%
\section{Conclusion}
We have shown that self-tuning models may solve two CC problems, one the fine-tuning problem and the other the huge inflationary vacuum energy problem. The solution of the fine-tuning problem needed an AdS bulk with an extra dimension(s) and an appropriate range of the brane tension, and the solution of the inflationary vacuum energy problem needed a solution region where the flat space is forbidden. Even though we used a specific example in deriving these features, they may survive in other self-tuning models.

\acknowledgments{I thank J.-H. Huh, B. Kyae and P. Yi for useful discussions. This work is supported in part by the Korea Research Foundation, Grant No. KRF-2005-084-C00001. }

%%%%%%%%%%%%%%%%%%%%%%%%%%%%%%%%%%%%%%%%%%%%%%%%%%%%%%%%%%%%%%%%%%%%%%%%%%%%
\vskip 0.5cm
\centerline{\bf Appendix: The AdS volume}

Here, we comment on our method of comparing the infinite volumes of the AdS spaces. The $n$-dimensional Euclidian space metric is given by
\dis{
ds^2 &=a^2\left(\frac{dr^2}{1-kr^2}+r^2 d\Omega^2_{n-1}\right)\\
&=a^2f^2(\eta)\left(d\eta^2+\eta^2 d\Omega^2_{n-1}\right)\label{eq:metric}
}
where $k=0,\pm 1$ and in the second equation the Weyl transformation is used since it is simple because of the vanishing Weyl tensor in this space,
\dis{
\frac{dr}{\sqrt{1-kr^2}}=f(\eta)d\eta,\quad f(\eta)=\frac{r}{\eta}~ .\label{eq:WeylTr}
}
Now, let us specify to the AdS space of $k=-1$. Then, Eq. (\ref{eq:WeylTr}) is integrated to give $\ln\eta=-\sinh^{-1} (1/r)$, or
\dis{
f(\eta)=\frac{r}{\eta}=\frac{2}{1-\eta^2}.
}
The Ricci scalar for the metric $g'_{\mu\nu}=a^2f^2(\eta)g_{\mu\nu}$ is given by
\dis{
R'= &a^{-2}f^{-2}\Big\{R-2(n-1)\nabla^2(\ln f)\\
&-(n-1)(n-2)\left(\frac{f'}{f}\right)^2\Big\}~ .\label{eq:Ricci}
}\\
Noting that $R'_{\mu\nu}=\overline{\Lambda}g_{\mu\nu}'$ or $R'=n\overline{\Lambda}$, we have $\overline{\Lambda}= -(n-1)/a^2$ in the
$n$-dimensional Euclidian AdS.

Using Eq. (\ref{eq:WeylTr}), the $n$-dimensional Euclidian AdS volume with the metric (\ref{eq:metric}) is regularized to
\dis{
V&_{{\rm AdS}^n} =a^n\int d^nx\left(\frac{2}{1-\eta^2}\right)^n
\\&=
a^n\int_0^1 d\eta ~\eta^{n-1}V_{S^{n-1}}\left(\frac{2}{1-\eta^2}\right)^n \\
&=(2a)^n V_{S^{n-1}}\frac12\int_0^1 d\xi ~\xi^{n/2-1}(1-\xi)^{-n}\\
&=\frac12(2a)^n V_{S^{n-1}} B(n/2, 1-n)\\
&=\frac12 (2a)^n\frac{2\pi^{n/2}}{\Gamma(n/2)} \frac{\Gamma(n/2)\Gamma(1-n)}{\Gamma(1-n/2)}\\
&=\left( \frac{4(n-1)\pi}{|\overline{\Lambda}|}\right)^{n/2}
\frac{\Gamma(1-n)}{\Gamma(1-n/2)}~ .\label{eq:AdSVol}
}
Eq. (\ref{eq:AdSVol}) factored out the diverging Gamma functions, and we can compare the $\overline{\Lambda}$ dependences. For $n=4$, it diverges as $1/|\overline{\Lambda}|^2$ as $\overline{\Lambda}$ tends to zero.

%%%%%%%%%%%%%%%%%%%%%%%%%%%%%%%%%%%%%%%%%%%%%%%%%%%%%%%%%%%%%%%%%%%%%%%%%%%

\end{document}